\documentclass[
    ,final            
  ]
  {aipproc}

\layoutstyle{8x11single}

\usepackage{amsfonts}

\begin{document}

\title{Tunnel determinants from spectral zeta functions. \\ Instanton effects in quantum mechanics}

\classification{11.15.Kc; 11.27.+d; 11.10.Gh}
\keywords      {Tunnel determinant, instanton, spectral zeta function}

\author{A. Alonso Izquierdo}{
  address={Departamento de Matematica
Aplicada and IUFFyM, Universidad de Salamanca, SPAIN}
}

\author{J. Mateos Guilarte}{
  address={Departamento de Fisica
Fundamental and IUFFyM, Universidad de Salamanca, SPAIN}
}

\begin{abstract}
In this paper we develop an spectral zeta function regularization procedure on the determinants of instanton fluctuation operators
that describe the semi-classical order of tunnel effects bewtween degenerate vacua.
\end{abstract}

\maketitle


\section{Introduction}
Tunneling phenomena between discrete degenerate vacua in quantum mechanics may be described at the semi-classical level as due to instanton paths in Euclidean time. The two main ingredients entering in the tunnel amplitudes are the exponential of the Euclidean instanton action and a quotient of determinants ressuming the fluctuations around the instanton over the classical vacuum fluctuations. Computation of this determinants have been achieved in the Literature frequently using variations of the Gelfand-Yaglom definition of determinants of differential operators,
see \cite{Gelfand}. Our purpose in this work is the application of the Ray-Singer determinants, see \cite{RaySinger}, to describe instanton physics in quantum mechanics.

Ray-Singer determinants of elliptic operators are defined from the derivatives of their spectral zeta functions.
Given the relationship between spectral zeta and heat functions via the Mellin's transform it is possible to gain information about the zeta function from the Gilkey-DeWitt heat kernel expansion, see \cite{Gilkey1984,DeWitt1965}. Application of these techniques in physics, specially effective dealing with one-loop effects in field theory either on flat or curved spaces, abound, see e.g. \cite{Elizalde1994,Kirsten2002,Vassilevich2003}. We plan to understand instanton effects in quantum mechanics within this framework in the hope that spectral zeta function analysis will deliver very good approximations even if the scattering data of the instanton fluctuation operators are unknown. To perform this attack to the problem we shall use the modified GDW expansion developed in \cite{Alonso2012a} and \cite{Alonso2013}
in order to tame the influence on zero mode fluctuations.

\section{Instantons in quantum mechanics}

Let us consider a one-dimensional system in quantum mechanics with dynamics generated by the Hamiltonian operator:
\begin{equation}
\hat{H}=-\frac{\hbar^2}{2}\frac{d^2}{dx^2}+U(x) \quad , \label{qham}
\end{equation}
where $x$ is the position of a particle of mass unit moving on the real line under the influence of the potential energy $U(x)$which is a bounded below function of the real line in $\mathbb{R}$. If $\tau \in (-\frac{T}{2},\frac{T}{2})$ denotes Euclidean time and
\begin{equation}
S_E[x]=\lim_{T \to\infty} \int_{-\frac{T}{2}}^\frac{T}{2}\, d\tau \, \left\{\frac{1}{2}\left(\frac{dx}{d\tau}\right)^2+U(x)\right\} \label{eac}
\end{equation}
refers to the Euclidean action, the Euclidean time evolution between two eigenstates of the position operator $\hat{x}\vert x\rangle=x\vert x\rangle $ is given by the following Feynman
path integral:
\begin{equation}
\langle x_f\vert {\rm exp}\Big[-\frac{T}{\hbar}\hat{H}\Big]\vert x_i \rangle=N\int\, {\cal D}[x(\tau)]\cdot{\rm exp}\Big[-\frac{1}{\hbar}S_E[x]\Big]=\sum_n \, {\rm exp}\Big[-\frac{T}{\hbar}E_n\Big]\cdot\langle x_f\vert n\rangle\langle n\vert x_i\rangle  \label{etev}.
\end{equation}
The Wiener integral on the central member of formula (\ref{etev}) is over the space of Euclidean trajectories respectively starting and ending at $x(-\frac{T}{2})=x_i$ and $x(\frac{T}{2})=x_f$ and $N$ is a normalization factor. The right-member expresses the amplitude in terms of the Hamiltonian eigenvalues and eigenfunctions: $\hat{H}\vert n\rangle=E_n\vert n\rangle$. It is clear that  for very large $T$ the ground state dominates the series.

We focus on potentials such that exhibit a discrete number of degenerate absolute minima:
\[
\frac{\delta U}{\delta x}\Big|_{x=x^{(a)}}=0 \quad , \quad \frac{\delta^2 U}{\delta x^2}|_{x=x^{(a)}}=m^2>0 \quad , \quad a=1,2,3, \dots \quad , \quad x^{(1)}<x^{(2)}<x^{(3)}<\cdots \, \, .
\]
Each minimum is a \lq\lq trivial\rq\rq{} Euclidean trajectory $\frac{d^2 x^{(a)}}{d\tau^2}= \frac{\delta U}{\delta x}|_{x=x^{(a)}\quad , \quad}=0$ where one expects on classical arguments to be centered the particle in one ground state. It may be, however, that the Euclidean action
admits finite Euclidean action instanton solutions among the Euclidean classical trajectories:
\begin{eqnarray*}
&& \frac{dx}{d\tau}=\sqrt{2 U(x)} \quad \equiv \quad \tau=\tau_0+\int \, \frac{dx}{\sqrt{2 U(x)}} \\
&& \lim_{\tau \to -\infty} \overline{x}(\tau)=x^{(a)} \, \, \, , \, \, \, \lim_{\tau\to \infty}\overline{x} (\tau)=x^{(a+1)} \, \, \,  ,
\quad S_E[\overline{x}]=\int_{x^{(a)}}^{x^{(a+1)}}
 \, dx \, \sqrt{2 U(x)}=S_0<\infty \, \, \, \, \, .
\end{eqnarray*}

\subsection{Tunnel effect through quantum mechanical instantons}

Tunnel effect between the classical minima $\vert x^{(a)}\rangle$ and  $\vert x^{(a+1)}\rangle$ obeys to the non null amplitude $\langle x^{(a+1)} \vert e^{-\frac{T}{\hbar}\hat{H}}\vert x^{(a)} \rangle$. The steepest descent method applied to the path integral formula in (\ref{etev}) runs over the one-instanton path as follows:

\noindent (1) One expands the Euclidean trajectories starting at $x^{(a)}$ and $x^{(a+1)}$ around the instanton,
\begin{equation}
x(\tau)=\overline{x}(\tau)+\sum_{n}\, c_n x_n(\tau) \, \, , \label{infl}
\end{equation}
where $x_n(\tau)$ form a family of orthonormal eigenfunctions of the Schr$\ddot{\rm o}$dinger operator
\begin{equation}
\mathbb{L}=-\frac{d^2}{d\tau^2}+v^2+V(\tau) \quad , \quad v^2+V(\tau)=\frac{d^2U}{dx^2}[\overline{x}(\tau)]\quad , \quad v^2= \frac{d^2U}{dx^2}[x^{(a)}]\quad , \quad \forall a
\end{equation}
with Dirichlet boundary conditions. Id est,
\[
\mathbb{L} x_n(\tau)=\lambda_n x_n(\tau) \quad , \quad x_n(\pm{\textstyle\frac{T}{2}})=0 \quad , \quad \int_{-\frac{T}{2}}^\frac{T}{2}\, x_n(\tau)x_m(\tau)=\delta_{nm} \quad .
\]
(2) In the $\hbar\to 0$ semi-classical range the integrand is concentrated around the saddle $\overline{x}(\tau)$ trajectory, the integration measure over the Euclidean paths is traded by integration in the expansion coefficients $c_n$, whereas the contributions of higher than quadratic fluctuations are negligible.

\noindent (3) Therefore, the one-instanton contribution to the tunnel effect amplitude reads:
\[
\langle x^{(a+1)} \vert e^{-\frac{T}{\hbar}\hat{H}}\vert x^{(a)} \rangle \simeq_{\hbar\to 0} N e^{-\frac{1}{\hbar}S_E(\overline{x})}\int_{-\infty}^\infty\, \prod_n dc_n \, e^{-\sum_n \, \lambda_n c_n^2 +{\cal O}(\hbar)}  = N e^{-\frac{1}{\hbar}S_E(\overline{x})} {\rm det} \, \,\mathbb{L}^{-\frac{1}{2}}\left(1+{\cal O}(\hbar)\right)
\]
The standard procedure of integrating over the instanton center $\tau_0$, instead of over the zero mode coefficient $c_0$, and the choice
of the infinite normalization factor as the determinant due to fluctuations around the equilibrium point
\[
N=\det\mathbb{L}_0^\frac{1}{2} \, \quad , \, \quad \mathbb{L}_0=-\frac{d^2}{dx^2}+v^2
\]
give the one-instanton contribution to the amplitude (\ref{etev}), see \cite{Polyakov,Coleman,Rubakov}:
\begin{equation}
K=\left(\frac{S_0}{2\pi\hbar}\right)^{\frac{1}{2}}\left|\frac{\det {\mathbb L}_0}{\det{\mathbb L}^\perp}\right|^\frac{1}{2} \qquad , \label{oidsa}
\end{equation}
where $\perp$ means that ${\mathbb L}$ acts only on the orthogonal subspace to the
instanton zero mode: $x_0(\tau)=\frac{1}{\sqrt{S_0}}\frac{d\bar x}{d\tau}$. We thus stress that the factor in (\ref{oidsa}) is the Jacobian of the change of integration coordinate to the instanton center, and that the zero mode contribution has been extracted form the determinant in the denominator.

\subsection{Spectral zeta functions and tunnel determinants}

The usual approach in instanton physics to the understanding of the differential operator determinants entering in formula (\ref{oidsa}) rely on the Gelfand-Yaglom procedure \cite{Gelfand} to deal with functional determinants, see also the recent comprehensive review by G. Dunne \cite{Dunne}. In this work, however, we choose to regularize the Schr$\ddot{\rm o}$dinger operator determinants in terms of their associated spectral zeta functions following the proposal of Ray and Singer in their seminal paper \cite{RaySinger}. Assuming a finite Euclidean time interval the spectra of both $\mathbb{L}$ and $\mathbb{L}_0$ are non-negative and discrete:
\[
\mathbb{L} x_n(\tau)=\omega_n^2 x_n(\tau) \quad , \quad \mathbb{L}_0 y_n(\tau)=\nu_n^2 y_n(\tau) \quad , \quad [\omega_n^2]=[\nu_n^2]=T^{-2}\, \, ,
\]
whereas the spectral zeta functions on the orthogonal subspace to the kernel of $\mathbb{L}$ are the series
\[
\zeta_{\mathbb{L}^\perp}(s)=\sum_{\omega_n>0}\,\frac{1}{\omega_n^{2 s}} \qquad , \qquad \zeta_{\mathbb{L}_0}(s)=\sum_{\nu_n}\,\frac{1}{\nu_n^{2 s}} \, \, ,
\]
which are not strictly convergent for values of the modulus of $s$ below some threshold. Nevertheless, it has been thoroughly studied that there exist analytic continuations of these spectral series to meromorhic functions in the complex parameter $s\in\mathbb{C}$-plane. Zeta function regularization is based on assigning the value of this function at a regular point as the sum of these divergent series arising in some physical problems, see \cite{Gilkey1984,Elizalde1994,Kirsten2002,Vassilevich2003}.
Therefore, we write:
\begin{equation}
\frac{\det {\mathbb L}_0}{\det {\mathbb L}^\perp}=\frac{\prod_{\nu_n} \, \nu_n^2}{\prod_{\omega_n>0}\, \omega_n^2}={\rm exp}\Big[\frac{d\zeta_{{\mathbb L}_0}}{ds}(0)-\frac{d\zeta^\perp_{\mathbb L}}{ds}(0)\Big]
\label{RSdet}
\end{equation}
as the regularized value of the quotient of the products of eigenvalues.

We are particularly interested  in very long intervals of Euclidean time such that we are led to deal with the tricky situation posed by the continuous spectrum. Computation of the spectral functions $\zeta_{{\mathbb L}^\perp}(s)$ and $\zeta_{{\mathbb L}_0}$ requires information about the spectra collected according the following procedure:

\begin{enumerate}

\item Consider the ${\mathbb L}x_\omega(x)=\omega^2x_\omega(x)$ spectral problem on a normalizaing interval $[-\frac{T}{2},\frac{T}{2}]$
of very large, but finite, Euclidean time $T$ with Dirichlet boundary conditions on the eigenfunctions as described in the Appendix of Reference \cite{Alonso2004}. We remark that, because the $V(\tau)$ potentials to be addressed are even under the Euclidean time reflection $\tau\to -\tau$ the eigenfunctions are either odd or even functions of $\tau$. Therefore, the usual set of odd functions complying with Dirichlet boundary conditions is enlarged by even functions which are non null at the origin.

\item Generically, in this class of problems there is one zero mode, $\ell$ bound states and a continuous spectrum
starting at the threshold $v^2$ in the spectrum of $\mathbb{L}$. The highest bound state may correspond to an eigenvalue either less than or equal to the scattering threshold: $\omega_\ell^2\leq v^2$. When equality is attained this bound state is buried at the continuous spectrum threshold and becomes a half-bound state: due to the one-dimensional Levinson theorem, a weight of $\frac{1}{2}$ must be assigned to it, see \cite{Barton}. The scattering eigenfunctions of ${\mathbb L}_0$ and ${\mathbb L}$ are characterized  by the following spectral densities:
\[
\rho_{{\mathbb L}_0}(\omega)=\frac{T}{2\pi} \quad , \quad \rho_{{\mathbb L}}(\omega)=\frac{T}{2\pi}+\frac{1}{2\pi}\frac{d\delta}{d\omega}(\omega) \quad .
\]
Here, $\delta(\omega)$ are the total phase shifts induced by $V(\tau)$.
\item The ${\mathbb L}_0$- and ${\mathbb L}^\perp$-heat traces are accordingly, see again the Reference in \cite{Alonso2004}:
\begin{eqnarray*}
h_{{\mathbb L}_0}(\beta)&=&\frac{1}{2}e^{-v^2\beta}+\frac{T}{\pi}\int_0^\infty \, d\omega \, e^{-(\omega^2+v^2)\beta}\\ h_{{\mathbb L}^\perp}(\beta)&=& \sum_{j=1}^{\ell-1}e^{-\omega_j^2\beta}+s_\ell e^{-\omega_\ell^2\beta}+2\int_0^\infty \, dk \,\rho_{{\mathbb L}}(\omega) e^{-(\omega^2+v^2)\beta} \quad ,
\end{eqnarray*}
where $s_\ell=\frac{1}{2}$ if $\omega_\ell^2=v^2$ and $s_\ell=1$ if $\omega_\ell^2<v^2$. $\beta$ is a fictitious inverse temperature with dimensions of $T^2$.

\item The Mellin transforms of the spectral heat traces are precisely the spectral zeta functions:
\[
\zeta_{{\mathbb L}_0}(s)= \frac{1}{\Gamma(s)}\int_0^\infty \, d\beta \, \beta^{s-1}h_{{\mathbb L}_0}(\beta)\, \, \quad , \, \, \quad \zeta_{{\mathbb L}^\perp}(s)= \frac{1}{\Gamma(s)}\int_0^\infty \, d\beta \, \beta^{s-1}h_{{\mathbb L}^\perp}(\beta)\quad .
\]
\end{enumerate}

\section{The quantum pendulum and the double well}
To test this procedure we shall compute the tunnel determinants in the two physical sytems more profusely dealt with in the Literature about instantons in quantum mechanics.
\subsection{The simple pendulum}
The Euclidean action is
\begin{equation}
S(z)=m\int d\tau \Big[\frac{l^2}{2}\left(\frac{dz}{d\tau}\right)^2+gl(1-\cos z)\Big] \nonumber
\end{equation}
where $l$ is the pendulum length and $z$ the angle that the rope form with the vertical. Clearly, the minima of $U(z)$ and the instanton trajectories are:
\begin{equation}
z^{(n)}=2\pi n \, \, , \, \, n\in\mathbb{Z} \quad , \quad \, \, \, \, \, \overline{z}(\tau)=4 \, {\rm arctan} \, e^{\Omega (\tau - \tau_0)} +2\pi n \nonumber
\end{equation}
where $\tau_0$ is the instanton center, $\Omega=\sqrt{\frac{g}{l}}$ is the pendulum frequency and the value of the instanton Euclidean action is: $S_0=8 m l^2\Omega $. Fluctuations around the instanton and the constant minima are determined from the operators:
\begin{eqnarray}
\mathbb{L}&=&-\frac{d^2}{d\tau^2}+ \Omega^2- \frac{2\Omega^2}{\cosh^2(\Omega\tau)} \quad , \quad \mathbb{L}_0=-\frac{d^2}{d\tau^2}+ \Omega^2 \label{pendif}\\
 v^2&=&\Omega^2  \qquad , \qquad  V(\tau)= -\frac{2\Omega^2}{\cosh^2(\Omega\tau)}  \nonumber \, .
\end{eqnarray}

In this case the instanton fluctuation operator ${\mathbb L}$ exhibits one zero mode as the unique bound state whereas the \lq\lq phase shifts\rq\rq{} and the difference of spectral densities with respect to that of $\mathbb{L}_0$ are:
\[
\delta(\omega)=2{\rm arctan}\frac{\Omega}{\omega} \quad , \quad \rho_{\mathbb L}(\omega)-\rho_{{\mathbb L}_0}(\omega)=-\frac{1}{\pi}\frac{\Omega}{\omega^2+\Omega^2}
\]
Both ${\mathbb L}$ and ${\mathbb L}_0$ have one half-bound state and, thus, their contributions to the spectral functions cancel upon subtraction. The \lq\lq renormalized\rq\rq{} heat and zeta functions are obtained in terms of the complementary Error function and Euler Gamma functions:
\[
h_{{\mathbb L}^\perp}(\beta)- h_{{\mathbb L}_0}(\beta)=-{\rm Erfc}[\Omega\sqrt{\beta}] \hspace{0.5cm},\hspace{0.5cm} \zeta_{{\mathbb L}^\perp}(s)- \zeta_{{\mathbb L}_0}(s) = -\frac{1}{\Omega^{2s}\sqrt{\pi}} \frac{\Gamma(s+\frac{1}{2})}{\Gamma(s+1)} \quad .
\]
The magic of the zeta function regularization procedure shows itself in the fact that at regular points in the $s$-complex plane $\zeta_{\mathbb L}(s)$ and $\zeta_{{\mathbb L}^\perp}(s)$ coincide. Even though the strict domains of convergence of the Mellin's transform of $h_{{\mathbb L}}(\beta)$ and $h_{{\mathbb L}^\perp}(\beta)$ are different their analytic continuations away the poles are the same. This means that the spectral zeta function are regularized to identical values counting or not the zero mode .

The derivatives of the zeta functions are needed in the definition of the regularized determinants{\footnote{$\gamma$ is the Euler constant, $\psi(z)=\frac{d\log\Gamma(z)}{dz}$ is the digamma function and $H_z$ denotes the harmonic number function.}}:
\[
D\zeta_{\mathbb{L}^\perp}(s)= \frac{d}{ds} [ \zeta_{{\mathbb L}^\perp}(s)- \zeta_{{\mathbb L}_0}(s) ]
= \frac{1}{\sqrt{\pi}}\frac{\Gamma(s+\frac{1}{2})}{\Gamma(s+1)} \left[H_s-H_{s-{\textstyle\frac{1}{2}}}\right]
\]
Only knowledge of the behaviour near zero is necessary to compute the quotient of determinants. From
\[
\zeta_{\mathbb{L}^\perp}(s) \simeq \gamma +
\psi({\textstyle\frac{1}{2}})-\log\Omega^2+{\cal O}(s) = \log\frac{1}{4\Omega^2}+{\cal O}(s)
\]
we obtain
\begin{equation}
\mathbf{D}=\frac{\det {\mathbb L}}{\det {\mathbb L}_0}={\rm exp} [ - D\zeta_{\mathbb{L}}(0)]=\frac{1}{4\Omega^2} \label{instpd}\, \, .
\end{equation}
Finally, the one-instanton contribution to the tunneling amplitude reads:
\begin{equation}
K=\left(\frac{S_0}{2\pi\hbar}\right)^\frac{1}{2}\cdot \mathbf{D}^{-\frac{1}{2}}=4\Omega\cdot \sqrt{\frac{m l^2\Omega}{\pi\hbar}}
\end{equation}
in perfect agreement with the result in the References \cite{Wipf,Dunne,Marcos}.

\subsection{The polynomic double well}
The Euclidean action is
\begin{equation}
S(x)=m\int d\tau \Big[\frac{1}{2}\left(\frac{dx}{d\tau}\right)^2+\frac{\lambda^2}{4}(x^2-a^2)^2\Big] \nonumber
\end{equation}
The minima of $U(x)$ and the instanton trajectories are:
\begin{equation}
x^{(\pm)}=\pm a \quad , \quad \, \, \, \, \, \overline{x}(\tau)=a \, {\rm tanh}\left(\Omega(\tau-\tau_0)\right) \nonumber
\end{equation}
where $\tau_0$ is the instanton center, $\Omega=\frac{a\lambda}{\sqrt{2}}$ is the oscillation frequency of the particle around both minima, and the value of the instanton Euclidean action is: $S_0=\frac{4}{3} m a^2\Omega $. Fluctuations around the instanton and the constant minima are determined from the operators:
\begin{eqnarray}
\mathbb{L}&=&-\frac{d^2}{d\tau^2}+ 4\Omega^2- \frac{6\Omega^2}{\cosh^2(\Omega\tau)} \quad , \quad \mathbb{L}_0=-\frac{d^2}{d\tau^2}+4 \Omega^2 \label{pendif2}\\
 v^2&=&4\Omega^2  \qquad , \qquad  V(\tau)= -\frac{6\Omega^2}{\cosh^2(\Omega\tau)}  \nonumber \, .
\end{eqnarray}

In the spectrum of ${\mathbb L}$ there are one zero mode, $\omega^2_0=0$, one bona fide bound state $\omega_1^2=3\Omega^2$, and one half-bound state: $\omega^2_2=4 \Omega^2$. The phase shifts of the continuous spectrum are:
\[
\delta(\omega)=-2 \arctan \frac{3 \Omega \omega}{2 \Omega^2-\omega^2} \quad ,
\]
whereas the spectral density after the subtraction of the spectral density of ${\mathbb L}_0$ reads:
\[
\rho_{\mathbb L}(\omega)-\rho_{{\mathbb L}_0}(\omega)=\frac{1}{2\pi}\frac{d\delta}{d\omega}=-\frac{\Omega}{\pi} \,\Big( \frac{1}{\Omega^2+\omega^2}+\frac{2}{4\Omega^2+\omega^2}\Big) \, \, .
\]
The \lq\lq renormalized\rq\rq{} ${\mathbb L}$ heat spectral function is accordingly:
\begin{eqnarray*}
h_{{\mathbb L}^\perp}(\beta)-h_{{\mathbb L}_0}(\beta)&=& e^{-3\Omega^2\beta}-\frac{2\Omega}{\pi}\cdot e^{-4\Omega^2\beta}\cdot\int_0^\infty\, d\omega\, \Big(\frac{1}{\omega^2+\Omega^2}+\frac{2}{\omega^2+4\Omega^2}\Big)e^{-\omega^2\beta}\\&=& e^{-3\Omega^2\beta}{\rm Erf}[\Omega\sqrt{\beta}]-{\rm Erfc}[2\Omega\sqrt{\beta}] \, \, .
\end{eqnarray*}
We remark that the contribution of the half-bound state of $\mathbb{L}$ has been canceled by the contribution of the half-bound state of $\mathbb{L}_0$. After this we calculate the difference between the spectral zeta functions via Mellin's transform:
\begin{eqnarray*}
\zeta_{{\mathbb L}^\perp}(s)-\zeta_{{\mathbb L}_0}(s)&=& \frac{\Gamma(s+\frac{1}{2})}{\sqrt{\pi}\Omega^{2s}\Gamma(s)} \left[
\frac{2}{3^{s+\frac{1}{2}}} \, {}_2F_1[{\textstyle \frac{1}{2},s+\frac{1}{2},\frac{3}{2};-\frac{1}{3}}]-\frac{1}{4^s} \frac{1}{s}\right]
\end{eqnarray*}
where ${}_2F_1[a,b,c,;z]$ is the Hypergeometric Gauss function. The difference between the zeta function derivatives needed in the regularization of the quotient of determinants looks very complicated:
\begin{eqnarray*}
\frac{d\zeta_{{\mathbb L}^\perp}}{ds}(s)-\frac{d\zeta_{{\mathbb L}_0}}{ds}(s) &=& \frac{1}{\Omega^{2s}} \frac{\Gamma[s+\frac{1}{2}]}{\sqrt{\pi}\Gamma[s]} \times \\ &\times&
\left[-2\cdot 3^{-\frac{1}{2}-s} {}_2F_1[{\textstyle \frac{1}{2},\frac{1}{2}+s,\frac{3}{2};-\frac{1}{3}}] \left(\log 3+\psi(s)-\psi({\textstyle\frac{1}{2}}+s) \right) + \frac{4^{-s}}{s^2} + \right. \\ &+& \left. \frac{4^{-2}}{s} \left( \log 4 + \psi(s) - \psi(s+{\textstyle\frac{1}{2}})\right) + 2\cdot 3^{-\frac{1}{2}-s} {}_2F_1^{(0,1,0,0)}[{\textstyle\frac{1}{2},\frac{1}{2}+s,\frac{3}{2};-\frac{1}{3}}]
\right]
\end{eqnarray*}
where $\, _2F_1^{(0,1,0,0)}[a,b,c;z]$ denotes the derivative of the Gauss Hypergeometric function with respect to the second argument. The Taylor expansion near the origin $s=0$
\[
\frac{d\zeta_{{\mathbb L}^\perp}}{ds}(\varepsilon)-\frac{d\zeta_{{\mathbb L}_0}}{ds}(\varepsilon)\simeq_{\varepsilon\to 0} -\gamma-\psi(\frac{1}{2})+\log (4\Omega^2)+2{\rm arcsinh}\frac{1}{\sqrt{3}}+{\cal O}(\varepsilon) = \log 48 \Omega^2+{\cal O}(\varepsilon)
\]
tells us that the quotient of determinants is:
\begin{equation}
\mathbf{D}=\frac{{\rm det}\,{\mathbb L}^\perp}{{\rm det}\,{\mathbb L}_0}={\rm exp}\Big[\frac{d\zeta_{{\mathbb L}_0}}{ds}(0)-\frac{d\zeta_{\mathbb L}}{ds}(0)\Big]=\frac{1}{48\Omega^2} \label{instdw} \quad .
\end{equation}
Therefore, the one-instanton contribution to the tunneling amplitude in the double well potential reads:
\begin{equation}
K=\left(\frac{S_0}{2\pi\hbar}\right)^\frac{1}{2}\cdot \mathbf{D}^{-\frac{1}{2}}=8\Omega\cdot \sqrt{\frac{m a^2\Omega}{2\pi\hbar}}
\end{equation}
now in perfect agreement with the result in \cite{Dunne,Marcos}.

\section{The heat trace asymptotics and zero modes}
The instanton fluctuation operators in these two physical systems are rather peculiar because are the first two members of the transparent P$\ddot{\rm o}$sch-Teller hierarchy and the whole spectral data are analytically known. The use of the zeta function procedure for other instanton models with more complicated spectral problems requires some indirect strategy like, for instance, use of high-temperature heat trace asymptotics. Moreover, because instanton tunneling always brings with it a zero mode we shall adapt to these problems the machinery developed
in the References \cite{Alonso2012a} and \cite{Alonso2013} to compute one-loop shifts to classical masses of topological kinks, an scenario where
zero modes also plays a r$\hat{\rm o}$le. We shall follow in this paper the heat kernel/zeta function regularization techniques on one-loop topological defect fluctuations summarized in the reviews \cite{Mateos2006,Mateos2009}

\subsection{Heat kernel asymptotic expansion and zero modes}

The problem is the formulation of the high-temperature heat kernel expansion in such a way that low-temperature effects, always important
when there are zero modes, are also incorporated. In this Section we shall describe a modification of the Gilkey-DeWitt heat kernel asymptotic expansion designed to cope with this problem. In particular, we shall focus on the ${\mathbb L}$-heat equation kernels associated to the instanton fluctuation operators of the form described in previous Sections:
\begin{eqnarray}
&& {\mathbb L} =-\frac{\partial^2}{\partial\tau^2}+v^2+V(\tau) \, \, \, , \, \, \, \tau\in \mathbb{R} \quad , \quad \lim_{\tau\rightarrow \pm \infty} V(\tau) =0
\label{operatorK} \\
&& \Big(\frac{\partial}{\partial\beta}+{\mathbb L}(\tau_1)\Big)K(\tau_1,\tau_2;\beta)=0 \quad , \quad K(\tau_1,\tau_2;0)=\delta(\tau_1-\tau_2) \label{asympintkernelht} \, \, \, .
\end{eqnarray}
We recall that the ${\mathbb L}$-spectrum
\[
{\rm Spec}({\mathbb L})=\{\omega_0^2 = 0\} \cup \{\omega_n^2\}_{n=1,\dots,\ell} \cup \{\omega^2 + v^2\}_{\omega\in \mathbb{R}}
\]
embraces a zero mode $x_0(\tau)$, $\ell$ bound states $x_n(\tau)$ with non-negative eigenvalues and scattering states $x_\omega(\tau)$ emerging on the threshold value $v^2$. The heat integral kernel is accordingly:
\begin{equation}
K_{{\mathbb L}}(\tau_1,\tau_2;\beta)= x_0(\tau_2) x_0(\tau_1)+\sum_{n=1}^\ell x_n(\tau_2)x_n(\tau_1) e^{-\beta \omega_n^2}+ \int \! [d\omega] \, x_\omega (\tau_2) \, x_\omega(\tau_1) \, e^{-\beta (\omega^2+v^2)} \quad .
\label{integralkernel}
\end{equation}
From this expression (\ref{integralkernel}) we deduce the behaviour of the heat integral kernel at low and high temperatures:
\begin{equation}
\lim_{\beta\rightarrow 0} K_{\mathbb L}(\tau_1,\tau_2;\beta)= \delta(\tau_1-\tau_2) \quad , \quad \lim_{\beta\rightarrow +\infty} K_{\mathbb L}(\tau_1,\tau_2;\beta)= x_0(\tau_2) x_0(\tau_1) \quad .
\label{asympintkernellt}
\end{equation}
At high temperature the condition (\ref{asympintkernelht}) is guaranteed due to the completeness of the set of eigenfunctions of ${\mathbb L}$
and it is independent of the existence of zero modes. The contribution of the zero mode, however, survives in the low temperature $\beta\to +\infty$ regime that captures the infrared part of the spectrum. Operators with trivial algebraic kernels (without zero modes) give rise to heat integral kernels that vanish at low temperatures under the assumption of positiveness  on  the ${\mathbb L}$-spectrum.
The heat integral kernel for the Helmholtz operator
\begin{equation}
{\mathbb L}_0=-\frac{d^2}{d\tau^2}+v^2\hspace{1cm}, \hspace{1cm} \tau\in \mathbb{R} \hspace{0.5cm}, \label{operatorK00}
\end{equation}
behaves exactly this way: ${\rm Spec}({\mathbb L}_0)=\{\omega^2+v^2\}_{\omega\in \mathbb{R}}$ such that  only presents scattering states, which are the $T\to +\infty$ limit of the even and odd eigenfunctions satisfying Dirichlet boundary conditions. From  $x_\omega^{0+}(\tau)=\frac{1}{\sqrt{2\pi}} {\rm cos}\omega\tau$ and $x_\omega^{0-}(\tau)=\frac{1}{\sqrt{2\pi}} {\rm sin}\omega\tau$ one finds:
\begin{eqnarray}
K_{{\mathbb L}_0}(\tau_1,\tau_2;\beta)&=&e^{-\beta v^2}\int_0^\infty \, d\omega \,e^{-\beta\omega^2} \left(x_\omega^{0+}(\tau_1)x_\omega^{0+}(\tau_2)+x_\omega^{0-}(\tau_1)x_\omega^{0-}(\tau_2)\right)=\nonumber \\ &=& \frac{1}{\sqrt{4\pi \beta\, }} \,
e^{-\beta v^2}\, e^{- \frac{(\tau_1-\tau_2)^2}{4 \, \beta\,}} \hspace{0.4cm},
\label{heatkernel0}
\end{eqnarray}
whose asymptotic behaviour respectively at high and low temperature is
\begin{equation}
\lim_{\beta\rightarrow 0} K_{{\mathbb L}_0}(\tau_1,\tau_2;\beta)= \delta(\tau_1-\tau_2) \hspace{0.5cm},\hspace{0.5cm}  \lim_{\beta\rightarrow +\infty} K_{{\mathbb L}_0}(\tau_1,\tau_2;\beta)= 0 \hspace{0.5cm}.
\label{asympintkernel0}
\end{equation}
The Gilkey-DeWitt ${\mathbb L}$-heat kernel $K_{\mathbb L}(\tau_1,\tau_2,\beta)$ search starts from  the standard factorization
\begin{equation}
K_{{\mathbb L}}(\tau_1,\tau_2;\beta)=K_{{\mathbb L}_0}(\tau_1,\tau_2;\beta) \, A(\tau_1,\tau_2;\beta) \quad  \mbox{with} \quad A(\tau_1,\tau_2;0)=1
\label{factorization0}
\end{equation}
taking profit of the precise analytical information about the ${\mathbb L}_0$-heat kernel given in (\ref{heatkernel0}).
If there is one zero mode the left and right members in the equality (\ref{factorization0}) differ beyond certain value $\beta_0$ when $\beta\rightarrow \infty$. Because we plan to use the heat trace expansion when the instanton fluctuation operators do not permit analytical
determination of the spectral data such that the spectral zeta functions are obtained via Mellin's transform over the whole $\beta$ range
it is necessary to modify the GDW factorization (\ref{factorization0}) to improve the effectiveness of the method.

\subsection{The modified Gilkey-DeWitt heat kernel expansion}

The idea is to start from the modified factorization
\begin{equation}
K_{{\mathbb L}}(\tau_1,\tau_2;\beta)=K_{{\mathbb L}_0}(\tau_1,\tau_2;\beta) \, C(\tau_1,\tau_2;\beta) +g(\beta) e^{-\frac{(\tau_1-\tau_2)^2}{4\beta}} x_0(\tau_2) x_0(\tau_1)\, \, \, ,
\label{factorization}
\end{equation}
 where the function $g(\beta)$ must be chosen such that the right member in (\ref{factorization}) fit in the asymptotic high and low temperature
 behaviours identified in (\ref{asympintkernellt}):
\begin{equation}
\lim_{\beta \rightarrow 0}C(\tau_1,\tau_2;0)=1 \hspace{1cm} , \hspace{1cm} \lim_{\beta \rightarrow \infty} g(\beta)=1 \hspace{0.4cm} , \hspace{0.4cm} \lim_{\beta \rightarrow 0}g(\beta)=0  \,\,\, . \label{heatic2}
\end{equation}
Therefore, the difference between the ${\mathbb L}$ and ${\mathbb L}_0$ heat traces reads:
\begin{eqnarray*}
h_{\mathbb L}(\beta)-h_{{\mathbb L}_0}(\beta) &=& \int_{-\infty}^\infty d\tau_1 K_{\mathbb L}(\tau_1,\tau_1;\beta) - \int_{-\infty}^\infty d\tau_1 K_{{\mathbb L}_0}(\tau_1,\tau_1;\beta) = \\ &=& \frac{e^{-\beta v^2}}{\sqrt{4\pi \beta}} \int_{-\infty}^\infty d\tau_1 \lim_{\tau_2\rightarrow \tau_1} [C(\tau_1,\tau_2;\beta)-1] + g(\beta)
\end{eqnarray*}
Plugging the ansatz (\ref{factorization}) into the ${\mathbb L}$-heat equation in (\ref{asympintkernelht}) we obtain the following \lq\lq transfer" equation for $C(\tau_1,\tau_2;\beta)$:
\begin{eqnarray}
\label{heateq2.all}
&&0=\left( \frac{\partial}{\partial \beta}+\frac{\tau_1-\tau_2}{\beta}
\frac{\partial}{\partial \tau_1}-\frac{\partial^2}{\partial
\tau_1^2}+V(\tau_1) \right) C(\tau_1,\tau_2;\beta) +  \label{heateq2.a} \\ && + \sqrt{4\pi \beta} \, e^{\beta v^2}\, x_0(\tau_2) \left[ \frac{dg(\beta)}{d\beta} x_0(\tau_1)+ \frac{g(\beta)}{2\beta} x_0(\tau_1) +\frac{g(\beta)}{\beta} (\tau_1-\tau_2) \frac{dx_0(\tau_1)}{d\tau_1} \right] \label{heateq2.b} \quad .
\end{eqnarray}
The terms specified in the line (\ref{heateq2.b}) should capture the zero mode contribution by means an appropriate choice of $g(\beta)$ and complement those solving the standard GDW \lq\lq transfer" equation (written in the (\ref{heateq2.a}) line above). We then try a power series expansion in $\beta$ to solve the PDE (\ref{heateq2.all}):
\begin{equation}
C(\tau_1,\tau_2;\beta) = \sum_{n=0}^\infty c_n(\tau_1,\tau_2) \, \beta^n  \quad \mbox{with} \quad c_0(\tau_1,\tau_2)=1 \quad .
\label{expansion}
\end{equation}
Note that the constraint $c_0(\tau_1,\tau_2)=1$ is forced by the first condition in (\ref{heatic2}). Plugging the modified factorization (\ref{expansion}) into (\ref{heateq2.all}) we find the infinite set of PDE's between the coefficients and the zero mode modification:
\begin{eqnarray}
& & \sum_{n=0}^\infty \left[ (n+1) c_{n+1}(\tau_1,\tau_2) - \frac{\partial^2 c_n(\tau_1,\tau_2)}{\partial \tau_1^2} +(\tau_1-\tau_2) \frac{\partial c_{n+1}(\tau_1,\tau_2)}{\partial \tau_1} +V(\tau_1) c_n(\tau_1,\tau_2) \right] \beta^n + \nonumber \\
& &  \hspace{0.5cm} +\sqrt{4\pi \beta} e^{\beta v^2} x_0(\tau_2) \left[ \frac{dg(\beta)}{d\beta} x_0(\tau_1) + \frac{g(\beta)}{2\beta} x_0(\tau_1) + (\tau_1-\tau_2) \frac{g(\beta)}{\beta} \frac{dx_0(\tau_1)}{d\tau_1} \right] = 0  \label{relation01} \quad .
\end{eqnarray}

\subsection{Optimum choice of $g(\beta)$}
There are many functions $g(\beta)$ complying with the asymptotics (\ref{heatic2}).
Because $g(\beta)$ will be understood as part of a heat trace we propose the following choice:
\begin{equation}
g(\beta)= e^{-v^2 \beta} \sum_{r=1}^N e^{\frac{r^2 v^2}{N^2}\beta} \, {\rm Erf}\, \Big(\frac{r v}{N} \sqrt{\beta} \, \Big)  \,\,\, .
\label{functiong}
\end{equation}
The reason is that $g(\beta)$, in the form (\ref{functiong}), arises as the heat trace for the reflectionless P\"oschl-Teller type operator
\[
{\mathbb L}_N=-\frac{d^2}{d\tau^2} + v^2 - \frac{(N+1)v^2}{N} \, {\rm sech}^2 \frac{v \tau}{N} \quad .
\]
The ${\mathbb L}_N$ operators have the same scattering threshold as ${\mathbb L}$ and, like ${\mathbb L}$, all of them possess a zero mode. Besides the functions (\ref{functiong}) comply with the asymptotic conditions (\ref{heatic2}) for any $N$, thus satisfying the requirements demanded to the function $g(\beta)$. The positive integer $N$ is free a priori. We shall see that in concrete physical instanton models there is an optimum value of $N$ to achieve a better approximation to the spectral functions.

Use of the series expansion of the Error function
\[
{\rm Erf}\, z = \frac{2}{\sqrt{\pi}} e^{-z^2} \sum_{n=0}^\infty \frac{2^n}{(2n+1)!!}z^{2n+1}
\]
in the PDE's (\ref{relation01}) leads to the recurrence relations between the coefficients $c_n(\tau_1,\tau_2)$ and their derivatives:
\begin{eqnarray}\label{recursive1.all}
&& \hspace{-1cm} 0=(n+1) \, c_{n+1}(\tau_1,\tau_2)-\frac{\partial^2 c_n(\tau_1,\tau_2)}{\partial
\tau_1^2} +(\tau_1-\tau_2) \frac{\partial c_{n+1}(\tau_1,\tau_2)}{\partial
\tau_1}+V(\tau_1) c_n(\tau_1,\tau_2) + \nonumber \\ && \hspace{-0.8cm} +x_0(\tau_2) x_0(\tau_1) 2v N H_{N,1} \delta_{n0} + (\tau_1-\tau_2) x_0(\tau_2) \frac{dx_0(\tau_1)}{d\tau_1} \frac{2^{n+2}}{(2n+1)!!} \Big(\frac{v}{N}\Big)^{2n+1} H_{N,-1-2n} +   \label{recursive1} \\ &&
\hspace{-0.8cm} + x_0(\tau_2)x_0(\tau_1) 2^{n+1} v^{2n+1} \left[ \left( \frac{1}{(2n-1)!!} + \frac{1}{(2n+1)!!}\right) \frac{H_{N,-1-2n}}{N^{2n+1}} - \frac{1}{(2n-1)!!} \frac{H_{N,1-2n}}{N^{2n-1}} \right] \nonumber
\end{eqnarray}
that generalize the usual recurrence relations arising in the usual GDW procedure. In the previous formula the generalized Harmonic numbers
enter: $H_{n,m}=\sum_{k=1}^n k^{-m}$.

\subsection{The modified GDW heat trace expansion}

The heat kernel \lq\lq diagonal\rq\rq{} is subsequently written in terms of the diagonal densities $c_n(\tau_1,\tau_1)=\lim_{\tau_2\rightarrow \tau_1} c_n(\tau_1,\tau_2)$  in (\ref{expansion}):
\begin{equation}
K_{\mathbb L}(\tau_1,\tau_1,\beta)=\lim_{\tau_2\rightarrow \tau_1} K_{\mathbb L}(\tau_1,\tau_2,\beta)= \frac{e^{-\beta v^2
}}{\sqrt{4 \pi \beta\,}} \sum_{n=0}^\infty c_n(\tau_1,\tau_1) \,
\beta^n + g(\beta) x_0(\tau_1) x_0(\tau_1) \hspace{0.3cm},
\label{factorization7}
\end{equation}
whereas the spectral ${\mathbb L}$-heat trace reads:
\begin{equation}
h_{\mathbb L}(\beta)={\rm Tr}_{L^2}\, e^{-\beta{\mathbb L}}=\lim_{T\to +\infty}\int_{-\frac{T}{2}}^\frac{T}{2} \, d\tau_1 \, K_{\mathbb L}(\tau_1,\tau_1;\beta)  \hspace{0.3cm}. \label{heatkernel6}
\end{equation}

The recurrence relations (\ref{recursive1.all}) should be sufficient to identify the diagonal densities but performing this calculation a sublety arises that requires to be dealt with care. Two kind of operations are involved: on one hand, the $\tau_2\rightarrow \tau_1$ limit must be attained. On the other hand, derivatives with respect to $\tau_1$ enter in formula (\ref{recursive1.all}). These two operations do not commute
and to cope with this problem it is convenient the introduction of a bit of additional notation:
\begin{equation}
{^{(k)}C}_n(\tau_1)=\lim_{\tau_2 \rightarrow \tau_1} \frac{\partial^k
c_n(\tau_1,\tau_2)}{\partial \tau_1^k} \quad , \quad {^{(k)} C}_0(\tau_1)=\lim_{\tau_2\rightarrow \tau_1} \frac{\partial^k
c_0}{\partial \tau_1^k}= \delta^{k0} \quad , \quad k=0,1,2,3, \cdots \, \, .
\label{newcoef}
\end{equation}
For $k=0$ the diagonal densities are described but the notation when $k=1,2,3, \cdots$ indicates that the right order is: first, take derivatives with respect to $\tau_1$, second, go to the diagonal limit. The behaviour of $^{(k)}C_0(\tau_1)$ is fixed in this formula in such a way that the high temperature asymptotics (\ref{expansion}) is ensured.

Taking the $k$-th derivative of (\ref{recursive1.all}) with respect to $\tau_1$ and passing later to the $\tau_2\to \tau_1$ limit, secondary recurrence relations between the densities ${^{(k)} C}_n(\tau_1)$ are obtained:
\begin{eqnarray}
{^{(k)} C}_n(\tau_1)& =&\frac{1}{n+k} \left[ \rule{0cm}{0.6cm} \right.
{^{(k+2)} C}_{n-1}(\tau_1) - \sum_{j=0}^k {k \choose j}
\frac{\partial^j V}{\partial \tau_1^j}\, \, {^{(k-j)}
C}_{n-1}(\tau_1) - \nonumber \\ &&  - x_0(\tau_1) \frac{d^k x_0(\tau_1)}{d\tau_1^k} \Big\{ 2 v N H_{N,1} \delta_{n-1,0} + k \frac{2^{n+1}}{(2n-1)!!} \Big(\frac{v}{N}\Big)^{2n-1} H_{N,1-2n} +\nonumber \\ && + 2^n v^{2n-1} \Big[ \Big( \frac{1}{(2n-3)!!} + \frac{1}{(2n-1)!!} \Big) \frac{H_{N,1-2n}}{N^{2n-1}} - \frac{1}{(2n-3)!!} \frac{H_{N,3-2n}}{N^{2n-3}} \Big] \Big\}
\left. \rule{0cm}{0.6cm} \right] \quad ,
\label{capitalAcoefficients}
\end{eqnarray}
which must be solved in descending order until the \lq\lq Seeley\rq\rq{} densities in (\ref{factorization7}) $c_n(\tau_1,\tau_1)={^{(0)}C}_n(\tau_1)$
are determined. For instance, the first two densities $c_n(\tau_1,\tau_1)$ derived from (\ref{capitalAcoefficients}) for the ${\mathbb L}$-operator are listed below:
\begin{eqnarray}
c_0(\tau_1,\tau_1)={^{(0)} C}_0(\tau_1)&=&1\,\,, \nonumber \\
c_1(\tau_1,\tau_1)={^{(0)} C}_1(\tau_1)&=&-V(\tau_1) -2(N+1)v x_0^2(\tau_1)\,\,, \label{Seley1}
\end{eqnarray}
where we observe that the zeroth order density is identical to the density arising in the ususal GDW heat kernel expansion, but zero modes start to modify the diagonal densities already at next order in the high temperature expansion.

Integration of these diagonal densities over the whole Euclidean time interval
\[
c_n({\mathbb L})=\lim_{T\to +\infty}\int_{-\frac{T}{2}}^\frac{T}{2}  d\tau_1 \, c_n(\tau_1,\tau_1)
\]
supply the Seeley coefficients $c_n({\mathbb L})$ to be used in (\ref{factorization7}) and (\ref{heatkernel6}). Finally,
we obtain the heat trace expansion
\[
h_{{\mathbb L}}(\beta) =\frac{e^{-\beta v^2
}}{\sqrt{4 \pi \beta\,}} \sum_{n=0}^\infty c_n({\mathbb L}) \,
\beta^n + e^{-\beta v^2} \sum_{r=1}^N e^{\frac{r^2v^2}{N^2}\beta} \,{\rm Erf}\, \Big(\frac{rv}{N}\sqrt{\beta}\Big)
\]
built from this modification of the GDW procedure. Assuming that the zero mode eigenfunction is normalized in $\lim_{T\to +\infty}[-\frac{T}{2},\frac{T}{2}]$, the first two Seeley coefficients are:
\begin{eqnarray}
c_0({\mathbb L})&=& \lim_{T\to +\infty} T\,\,, \nonumber\\
c_1({\mathbb L})&=&- \left< V\right>  -2 (N+1) v\,\,, \quad \left< V\right>=\lim_{T\to +\infty} \int_{-\frac{T}{2}}^\frac{T}{2}  d\tau_1 \,V(\tau_1) \label{Seley2}
\end{eqnarray}
The linear divergence arising in $c_0({\mathbb L})$ is tammed by subtracting the ${\mathbb L}_0$-heat trace, renormalization that  amounts to dropping the $c_0({\mathbb L})$ coefficient:
\begin{equation}
h_{{\mathbb L}}(\beta)-h_{{\mathbb L}_0}(\beta) = \frac{e^{-\beta v^2
}}{\sqrt{4 \pi}} \sum_{n=1}^\infty c_n({\mathbb L}) \,
\beta^{n-\frac{1}{2}} +  e^{-\beta v^2} \sum_{r=1}^N e^{\frac{r^2 v^2}{N^2}\beta} {\rm Erf}\,\Big(\frac{rv}{N}\sqrt{\beta}\Big)  \,\,\, .
\label{heatfunction3}
\end{equation}

\section{Tunnel determinants from the heat trace expansion}

Mellin's transform of the renormalized heat trace expansion leads to the following expression of the difference between spectral zeta functions:
\begin{eqnarray}
&& \hspace{-1cm}\zeta_{\mathbb{L}}(s)-\zeta_{{\mathbb L}_0}(s)=
\frac{1}{\sqrt{4\pi} \Gamma[s]} \sum_{n=1}^\infty c_n(\mathbb{L}) v^{1-2n-2s} \Gamma[s+n-{\textstyle\frac{1}{2}}] + \nonumber\\ && + \frac{2v^{-2s}}{\sqrt{\pi} N} \frac{\Gamma[s+\frac{1}{2}]}{\Gamma[s]} \sum_{r=1}^{N-1} r \Big(1-\frac{r^2}{N^2}\Big)^{-\frac{1}{2}-s} {}_2F_1[{\textstyle\frac{1}{2},\frac{1}{2}+s,\frac{3}{2},\frac{r^2}{r^2-N^2}}] - \frac{v^{-2s}\Gamma[s+\frac{1}{2}]}{\sqrt{\pi} s \Gamma[s]}
\, \, . \label{mgdwsz}
\end{eqnarray}
Derivation of (\ref{mgdwsz}) with respect to $s$ reads
\begin{eqnarray*}
&& \hspace{-1cm} \frac{d\zeta_{\mathbb{L}}}{ds}(s)-\frac{d\zeta_{{\mathbb L}_0}}{ds}(s)= \frac{1}{\sqrt{4\pi} \Gamma[s]} \sum_{n=1}^\infty c_n(\mathbb{L}) v^{1-2n-2s} \Gamma[s+n-{\textstyle \frac{1}{2}}] \Big( \psi(s+n-{\textstyle\frac{1}{2}}) - \psi(s) -\log v^2 \Big) + \\
&& + \frac{2v^{-2s}}{\sqrt{\pi}N} \frac{\Gamma[s+\frac{1}{2}]}{\Gamma[s]} \sum_{r=1}^{N-1} r \Big( 1-\frac{r^2}{N^2} \Big)^{-\frac{1}{2}-s} \Big[ {}_2F_1^{(0,1,0,0)}[{\textstyle \frac{1}{2},\frac{1}{2}+s,\frac{3}{2},\frac{r^2}{r^2-N^2}}] - \\ &&   -   {}_2 F_1[{\textstyle\frac{1}{2},\frac{1}{2}+s, \frac{3}{2}, \frac{r^2}{r^2-N^2}}] \Big( \log(\textstyle{1-\frac{r^2}{N^2}}) + \log v^2 + \psi(s)-\psi (s+\frac{1}{2}) \Big) \Big] + \\ && + \frac{v^{-2s}}{\sqrt{\pi}} \frac{\Gamma[s+\frac{1}{2}]}{s^2 \Gamma[s]} \Big( 1+ 2s \log v +s \psi(s) - s \psi(s+{\textstyle \frac{1}{2}} )\Big)  \, \, ,
 \end{eqnarray*}
whereas the leading order of the Taylor expansion around $s=0$
\[
\frac{d\zeta_{\mathbb{L}}}{ds}(s)-\frac{d\zeta_{{\mathbb L}_0}}{ds}(s)\approx_{s\to 0}  \frac{1}{\sqrt{4\pi}} \sum_{n=1}^\infty c_n(\mathbb{L}) v^{1-2n} \Gamma[n-{\textstyle\frac{1}{2}}] + \log (4 v^2) + 2 \sum_{r=1}^{N-1} \, {\rm arcsinh} \, \frac{r}{\sqrt{N^2-r^2}} + {\cal O}(s)
\]
provides a complicated formula for the renormalized one-instanton determinant:
\begin{eqnarray*}
\mathbf{D}&=&\frac{\det {\mathbb L}}{\det {\mathbb L}_0}= \exp \Big[ \frac{d\zeta_{{\mathbb L}_0}}{ds}(0)  - \frac{d\zeta_{\mathbb{L}}}{ds}(0)\Big] = \\
&=&  \frac{1}{4 v^2} \exp \Big[-\frac{1}{\sqrt{4\pi}}\sum_{n=1}^{\infty} c_n(\mathbb{L})\, \frac{\Gamma(n-\frac{1}{2})}{v^{2n-1}} - 2 \sum_{r=1}^{N-1} \, {\rm arcsinh} \, \frac{r}{\sqrt{N^2-r^2}} \Big] \, \, \, .
\end{eqnarray*}
Use of the identity
\[
{\rm arcsinh}\frac{r}{\sqrt{N^2-r^2}}=\frac{1}{2}{\rm log}\frac{N+r}{\sqrt{N^2-r^2}}
\]
allows us to write the quotient of determinants in the more compact form:
\begin{equation}
\mathbf{D}= \frac{1}{4 v^2} \,\, \frac{N! (N-1)!}{(2N-1)!} \,\, \exp \Big[-\frac{1}{\sqrt{4\pi}}\sum_{n=1}^{\infty} c_n(\mathbb{L})\, \frac{\Gamma(n-\frac{1}{2})}{v^{2n-1}} \Big] \label{insthk} \quad .
\end{equation}
The main idea in this paper is to use the formula (\ref{insthk}) in instanton models where the spectral information about the ${\mathbb L}$
operator is insufficient to exactly evaluate $\zeta_{\mathbb L}(s)$ whereas the Seeley coefficients $c_n({\mathbb L})$ are computable.

\subsection{The double well again}
To test this procedure we shall compute the one-instanton determinant in the polynomic double well potential treated in  the subsection \S.3.2.
We write the instanton and unstable equilibrium point fluctuation operators with a re-scaling of Euclidean time $\Omega \tau\to\tau$ as follows:
\[
{\mathbb L}=-\frac{d^2}{d \tau^2} + 4 - \frac{6}{\cosh^2 \tau} \hspace{0.5cm},\hspace{0.5cm} {\mathbb L}_0=-\frac{d^2}{d \tau^2} + 4 \quad .
\]
The normalized zero mode eigenfunction is $x_0(\tau)=\frac{\sqrt{3}}{2}{\rm sech}^2\tau$ and in the next Table we list the Seeley coefficients
up to $c_{10}({\mathbb L})$ obtained by solving the recurrence relations above for the four lower choices of $g(\beta)$: $N=0,1,2,3$.

\centerline{\begin{tabular}{|c|cccc|} \hline
 & \multicolumn{4}{|c|}{Seeley coefficients} \\
$n$ & $N=0$ & $N=1$ & $N=2$ & $N=3$ \\ \hline
1 & 12.0000 & 4.00000 & 0.0 & -4.00000 \\
2 & 24.0000 & 2.66667 & 0.0 & -4.44440 \\
3 & 35.2000 & 1.06667 & 0.0 & -3.56872 \\
4 & 39.3143 & 0.304762 & 0.0 & -1.99621 \\
5 & 34.7429 & 0.0677249 & 0.0 & -0.836017 \\
6 & 25.2306 & 0.0123136 & 0.0 & -0.279378 \\
7 & 15.5208 & 0.0018944 & 0.0 & -0.0778552 \\
8 & 8.27702 & 0.000252587 & 0.0 & -0.0186493 \\
9 & 3.89498 & 0.0000297161 & 0.0 & -0.00392353 \\
10 & 1.63998 & $3.12801 \times 10^{-6}$ & 0.0 & -0.000736656 \\
11 & 0.624754 & $2.97906 \times 10^{-7}$ & 0.0 & -0.000124956 \\
12 & 0.217306 & $2.59049 \times 10^{-8}$ & 0.0 & -0.0000193371 \\ \hline
\end{tabular}}

\vspace{0.3cm}
The $N=0$ choice of $g(\beta)$ reproduces the ordinary GDW results without caring about the infrared contribution of the zero mode. $N=2$ is the optimum election vanishing all the Seeley coefficients. Therefore, we understand the meaning of $g(\beta)$ for $N=2$: it is precisely the exact heat trace of ${\mathbb L}_2$, the second P$\ddot{\rm o}$sch-Teller operator in the transparent hierarchy. If $g(\beta)$ with $N=1$ is used to
modify the GDW procedure the outcome is equivalent to substract from the ${\mathbb L}$-heat trace the ${\mathbb L}_1$-heat trace, in a certain sense the simplest operator with a zero mode. In physical language we could say that one measures the double well instanton fluctuations with respect to the instanton fluctuations in a simple pendulum. In Figure 1, one observes how the coefficients go very slowly to zero with increasing $n$ for $N=0$, a behaviour that announces non fast convergence properties in the series (\ref{mgdwsz}). The $N=1$ choice for $g(\beta)$ improves greatly on this situation showing that including infrared effects one needs less terms in (\ref{mgdwsz}) to achieve a good approximation to the heat trace. There is overshooting for $N=3$: use of $g(\beta)$ with $N=3$ gives rise to negative coefficients.

\begin{figure}[htdp]
\centerline{\includegraphics[height=5cm]{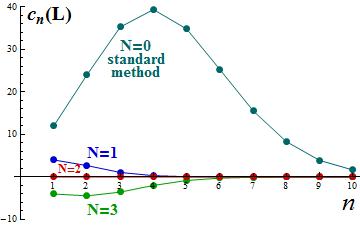}}
\caption{\small Dependence on $n$ of $c_n({\mathbb L})$ up to $n=10$ for the $N=0,1,2,3$ values in $g(\beta)$ }
\end{figure}

It is clear that the optimum choice for the instanton pendulum operator is $N=1$. The formula (\ref{insthk}) provides the very well known result:
$\mathbf{D}=\frac{1}{4\Omega^2}$ because in the pendulum case we have, if $N=1$, $v^2=\Omega^2$, and all the Seeley coefficients vanish $c_n(\mathbb{L})=0$. For identical reasons the one-instanton determinant in the double well is precisely reproduced by (\ref{insthk}): $\mathbf{D}=\frac{1}{48\Omega^2}$. We have chosen $N=2$ such that all the Seeley coefficients vanish $c_n(\mathbb{L})=0$ and for the double well we have $v^2=4\Omega^2$. It is interesting to consider what happens if another choices of $N$ are taken in this case. Use of the coefficients collected in the Table above provides, after truncation of the series in the exponent of formula (\ref{mgdwsz}) at the order tenth, the following results:
\begin{center}
\begin{tabular}{|c|c|c|c|} \hline
 & $N=1$ & $N=2$ & $N=3$ \\ \hline
$\mathbf{D}$ & $0.0208333 \Omega^{-2}$ & $\frac{1}{48}\Omega^{-2}$ & $0.0208330\Omega^{-2}$ \\ \hline
\end{tabular}
\end{center}
We see that taking $N=1$ is a very good option that gives an extremely accurate aproximation to the exact value.

\section{The Razavy and Khare double wells}

In the References \cite{Razavy} and \cite{Khare} it has been respectively proposed a double well potential based on hyperbolic functions and
a one-parametric family of sixth-order polynomic double wells that share the property of being quasi-exactly solvable systems. Both the Razavy and Khare potentials admit instanton solutions although the associated instanton operators are complicated enough to elude a direct calculation of the one-instanton determinants along the lines developed in Section \S.3. The heat trace expansion procedure is the only strategy allowing these calculations and the purpose of this Section is to perform this task.

\subsection{The Razavy potential}

The Euclidean action in the Razavy system is:
\begin{equation}
S[x]=m\int \, d\tau \, \Big\{\frac{1}{2}\left(\frac{dx}{d\tau}\right)^2+\frac{\alpha^2}{4}\Big({\rm sinh}^2[\frac{\Omega}{\alpha}x]-1\Big)^2\Big\} \, \, \, . \label{eucR}
\end{equation}
Like the polynomic double well $U(x)$ exhibits two minima, which are unstable points in the Euclidean mechanical system, There are also instanton trajectories:
\begin{equation}
x^{(\pm)}=\pm \frac{\alpha}{\Omega}{\rm arcsinh} 1 \quad , \quad \, \, \, \, \, \overline{x}(\tau)=\pm\frac{\alpha}{\Omega}{\rm arctanh}\Big[ \frac{1}{\sqrt{2}}\, {\rm tanh}\left(\Omega(\tau-\tau_0)\right)\Big] \, \, \, .\nonumber
\end{equation}
The instanton Euclidean action is:
\[
S_0=\Big[\frac{3\sqrt{2}}{2}{\rm arcsinh}1-1\Big] \frac{m\alpha^2}{\Omega}  \quad .
\]
Fluctuations around the instanton and the constant minima are determined from the operators:
\begin{eqnarray}
\mathbb{L}&=&-\frac{d^2}{d\tau^2}+ 4\Omega^2- \frac{4\Omega^2\left(11+9{\rm cosh}[2\Omega\tau]\right)}{\left(3+\cosh[2\Omega\tau]\right)^2} \quad , \quad \mathbb{L}_0=-\frac{d^2}{d\tau^2}+4 \Omega^2 \label{pendif3}\\
 v^2&=&4\Omega^2  \qquad , \qquad  V(\tau)= - \frac{4\Omega^2\left(11+9{\rm cosh}[2\Omega\tau]\right)}{\left(3+\cosh[2\Omega\tau]\right)^2}  \nonumber \, .
\end{eqnarray}
The instanton well in (\ref{pendif3}) is too complicated for unveiling the scattering data analytically. Only the zero mode is known:
\[
x_0(\tau)=\sqrt{\frac{m}{\Omega S_0}}\frac{2\sqrt{2}\alpha}{3+{\rm cosh}[2 \Omega \tau]} \quad .
\]
Therefore we rely on the zeta function formula (\ref{mgdwsz}) to compute the one-instanton determinant. The heat trace coefficients $c_n(\mathbb{L})$, which have been evaluated by solving the recurrence relations coming from the modified heat kernel expansion with different choices of function $g(\beta)$ for this instanton well, are listed in the next Table

\centerline{\begin{tabular}{|c|cccc|} \hline
 & \multicolumn{4}{|c|}{Seeley coefficients} \\
$n$ & $N=0$ & $N=1$ & $N=2$ & $N=3$ \\ \hline
1 & 15.4787 & 7.47870 & 3.47870 & -0.521297 \\
2 & 29.1604 & 7.82708 & 5.16041 & 0.715970 \\
3 & 39.8523 & 5.71900 & 4.65234 & 1.083616 \\
4 & 42.1618 & 3.15228 & 2.84751 & 0.851307 \\
5 & 36.0361 & 1.36104 & 1.29331 & 0.457301 \\
6 & 25.7003 & 0.482076 & 0.469763 & 0.190385 \\
7 & 15.6633 & 0.144365 & 0.142469 & 0.0647288 \\
8 & 8.314336 & 0.037568 & 0.0373159 & 0.0224673 \\
9 & 3.90348 & -0.0263172 & 0.00850317 & - \\
10 & 1.64181 & 0.0018300 & - & - \\\hline
\end{tabular}}

In Figure 2 one sees that the coefficients go rapidly to zero for the $N=1$, $N=2$, and $N=3$ cases, this last choice seems to be optimum.
\vspace{0.3cm}
\begin{figure}[htdp]
\centerline{\includegraphics[height=5cm]{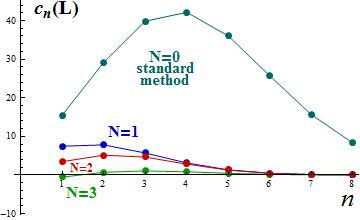}}
\caption{\small Seeley coefficients for the Razavy instanton up to the tenth order}
\end{figure}

The one-instanton determinant obtained by accounting up to ten coefficients in formula (\ref{mgdwsz}) for each $N=1,2,3$ option is:

\begin{center}
\begin{tabular}{|c|c|c|c|} \hline
 & $N=1$ & $N=2$ & $N=3$ \\ \hline
$\mathbf{D}$ & $0.0068021(\Omega^{2})^{-1}$ & $0.00679574(\Omega^{2})^{-1}$ & $0.0067951(\Omega^{2})^{-1}$ \\ \hline
\end{tabular}
\end{center}

\subsection{The family of Khare potentials}

The Euclidean action in the one-parametric family of Khare systems is:
\begin{equation}
S[x,a]=m\int \, d\tau \, \Big\{\frac{1}{2}\left(\frac{dx}{d\tau}\right)^2+\frac{\lambda^2}{2}\Big(x^2+ \frac{\Omega}{\lambda} a^2\Big) \Big(x^2-\frac{\Omega}{\lambda}\Big)^2\Big\} \, \, \, . \label{eucK}
\end{equation}
Again the energy potentials in the family $U(x,a)$ have two minima, which are unstable points in the Euclidean mechanical system and instanton trajectories interpolating between these unstable points:
\begin{equation}
x^{(\pm)}=\pm \sqrt{\frac{\Omega}{\lambda}} \quad , \quad \, \, \, \, \, \overline{x}(\tau,a)=\pm a \sqrt{\frac{\Omega}{\lambda}}\frac{{\rm tanh} [\Omega\sqrt{a^2+1}(\tau-\tau_0)]}{\sqrt{a^2+\,{\rm sech}^2 [\Omega\sqrt{a^2+1}(\tau-\tau_0)]}}\, \, \, .\nonumber
\end{equation}
The instanton Euclidean action is:
\[
S_0=\frac{\Omega^2 m a^2}{4 \lambda} \Big[ (a^2+4)\,{\rm arctanh} \Big(\frac{1}{\sqrt{a^2+1}}\Big)+2 -a^2\Big] \quad .
\]
Fluctuations around the instanton and the constant minima are determined from the operators:
\begin{eqnarray}
\mathbb{L}(a)&=&-\frac{d^2}{d\tau^2}+ 4\Omega^2 (1+a^2) \Big[ 1+ \frac{15(1+a^2)}{(2+a^2+a^2 \cosh (2 \Omega \sqrt{1+a^2} \tau))^2} - \frac{3(3+a^2)}{2+a^2+a^2 \cosh(2\Omega \sqrt{1+a^2} \tau)}\Big] \nonumber \\  \mathbb{L}_0(a)&=&-\frac{d^2}{d\tau^2}+4\Omega^2 (1+a^2) \quad , \quad v^2=4\Omega^2 (1+a^2) \label{pendif4}\\
 V(\tau,a)&=&4\Omega^2 (1+a^2) \Big[\frac{15(1+a^2)}{(2+a^2+a^2 \cosh (2 \Omega \sqrt{1+a^2} \tau))^2} - \frac{3(3+a^2)}{2+a^2+a^2 \cosh(2\Omega \sqrt{1+a^2} \tau)}\Big] \nonumber \, .
\end{eqnarray}
The instanton wells in (\ref{pendif4}) are again too complicated for obtaining the scattering data analytically. Only the zero modes are known:
\[
x_0(\tau,a)=\sqrt{\frac{m}{\Omega S_0}} \frac{a \Omega (a^2+1)^{\frac{3}{2}}\,{\rm sech}^2[\Omega \sqrt{a^2+1} \tau]}{\sqrt{\lambda}\left(a^2+\,{\rm sech}^2[\Omega\sqrt{a^2+1} \tau]\right)^\frac{3}{2}} \quad .
\]
The only way to estimate the one-instaton determinants is use of formula (\ref{mgdwsz}). The Seeley coefficients needed to feed this formula are collected in subsequent Tables for several values of the $a$ parameter.

{\small\centerline{\begin{tabular}{|c|ccc|} \hline
 & \multicolumn{3}{|c|}{Seeley coefficients $a=0.01$} \\
$n$ & $N=1$ & $N=2$ & $N=3$ \\ \hline
1 &  38.7888 & 34.7886 & 30.7884 \\
2 &  63.1003 & 60.4333 & 55.9881 \\
3 &  73.1756 & 72.1086 & 68.5390 \\
4 &  66.8426 & 66.5277 & 64.5408 \\
5 &  50.6752 & 50.6074 & 49.7711 \\
6 &  32.9228 & 32.9105 & 32.6310 \\
7 &  18.7091 & 18.7072 & 18.6293 \\
8 &  9.43570 & 9.43545 & 9.41678 \\
9 &  4.26962 & 4.26959 & 4.26566 \\
10 & 1.74936 & 1.74936 & 1.74862 \\
11 & 0.653606& 0.653606& 0.653481 \\\hline
\end{tabular}
\begin{tabular}{|c|ccc|} \hline
 & \multicolumn{3}{|c|}{Seeley coefficients $a=0.2$} \\
$n$ & $N=1$ & $N=2$ & $N=3$ \\ \hline
1 &  20.7358 & 16.6565 & 12.5773 \\
2 &  36.2191 & 33.3908 & 28.6771 \\
3 &  46.4620 & 45.2854 & 41.3491 \\
4 &  46.9750 & 46.6254 & 44.3355 \\
5 &  38.9032 & 38.8224 & 37.8250 \\
6 &  27.1512 & 27.1360 & 26.7893 \\
7 &  16.3108 & 16.3083 & 16.2079 \\
8 &  8.57958 & 8.57924 & 8.55421 \\
9 &  4.00591 & 4.00587 & 4.00039 \\
10 & 1.68009 & 1.68008 & 1.67901 \\
11 & 0.63851& 0.638571& 0.638382 \\\hline
\end{tabular}}}

\vspace{0.2cm}

{\small\centerline{\begin{tabular}{|c|ccc|} \hline
 & \multicolumn{3}{|c|}{Seeley coefficients $a=0.4$} \\
$n$ & $N=1$ & $N=2$ & $N=3$ \\ \hline
1 &  16.6319 & 12.3238 & 8.01568 \\
2 &  30.7569 & 27.4253 & 21.8726 \\
3 &  41.8698 & 40.3239 & 35.1519 \\
4 &  44.5980 & 44.0856 & 40.7297 \\
5 &  38.5230 & 38.3910 & 36.7606 \\
6 &  27.8046 & 27.7767 & 27.1447 \\
7 &  17.1601 & 17.1551 & 16.9508 \\
8 &  9.22966 & 9.22889 & 9.17212 \\
9 &  4.39096 & 4.39086 & 4.37700 \\
10 & 1.87271 & 1.87270 & 1.86968 \\
11 & 0.721927& 0.721925& 0.721332 \\\hline
\end{tabular}
\begin{tabular}{|c|ccc|} \hline
 & \multicolumn{3}{|c|}{Seeley coefficients $a=0.6$} \\
$n$ & $N=1$ & $N=2$ & $N=3$ \\ \hline
1 &  14.4796 & 9.81485 & 5.15008 \\
2 &  28.7922 & 24,5628 & 17.5138 \\
3 &  41.8878 & 39.5871 & 31.8894 \\
4 &  47.2700 & 46.3760 & 40.5201 \\
5 &  42.8619 & 42.5917 & 39.2564 \\
6 &  32.2567 & 32.1898 & 30.6740 \\
7 &  20.6529 & 20.6390 & 20.0645 \\
8 &  11.4905 & 11.4880 & 11.3008 \\
9 &  5.64176 & 5.64136 & 5.58781 \\
10 & 2.48460 & 2.48455 & 2.47087 \\
11 & 0.983229& 0.983222& 0.980067 \\\hline
\end{tabular}}}

\vspace{0.2cm}

{\small\centerline{\begin{tabular}{|c|ccc|} \hline
 & \multicolumn{3}{|c|}{Seeley coefficients $a=0.8$} \\
$n$ & $N=1$ & $N=2$ & $N=3$ \\ \hline
1 &  13.2386 & 8.11605 & 2.99355 \\
2 &  28.7991 & 23.1985 & 13.8642 \\
3 &  45.3404 & 41.6664 & 29.3744 \\
4 &  54.8504 & 53.1289 & 41.8528 \\
5 &  52.8468 & 52.2194 & 44.4746 \\
6 &  42.0251 & 41.8380 & 37.5934 \\
7 &  28.3195 & 28.2723 & 26.3324 \\
8 &  16.5776 & 16.5673 & 15.8052 \\
9 &  8.53925 & 8.53726 & 8.27432 \\
10 & 3.96706 & 3.96671 & 3.88575 \\
11 & 1.61838& 1.61832& 1.59580 \\\hline
\end{tabular}
\begin{tabular}{|c|ccc|} \hline
 & \multicolumn{3}{|c|}{Seeley coefficients $a=1.0$} \\
$n$ & $N=1$ & $N=2$ & $N=3$ \\ \hline
1 &  12.5436 & 6.88676 & 1.22991 \\
2 &  30.2796 & 22.7371 & 10.1663 \\
3 &  52.1899 & 46.1559 & 25.9681 \\
4 &  68.4586 & 65.0106 & 42.4261 \\
5 &  70.9337 & 69.4012 & 50.4843 \\
6 &  60.4224 & 59.8651 & 47.2219 \\
7 &  43.5063 & 43.3348 & 36.2882 \\
8 &  27.2918 & 27.2460 & 23.8702 \\
9 &  14.8911 & 14.8803 & 13.4598 \\
10 & 7.61304 & 7.61078 & 7.07738 \\
11 & 2.94085 & 2.09195 & 2.75946 \\\hline
\end{tabular}}}

\vspace{0.2cm}

{\small\centerline{\begin{tabular}{|c|ccc|} \hline
 & \multicolumn{3}{|c|}{Seeley coefficients $a=1.2$} \\
$n$ & $N=1$ & $N=2$ & $N=3$ \\ \hline
1 &  12.2086 & 5.96041 & -0.287793 \\
2 &  33.0300 & 22.8663 & 5.92674 \\
3 &  62.8828 & 52.9630 & 19.7745 \\
4 &  90.2650 & 83.3495 & 38.0523 \\
5 &  101.627 & 97.8776 & 51.5895 \\
6 &  93.8728 & 92.2093 & 54.4663 \\
7 &  73.2125 & 72.5880 & 46.9242 \\
8 &  49.9497 & 49.7466 & 34.7468 \\
9 &  28.8849 & 28.8266 & 21.1266 \\
10 & 17.9053 & 17.8904 & 14.3629 \\
11 & 2.47448 & 2.47103 & 1.01115 \\\hline
\end{tabular}
\begin{tabular}{|c|ccc|} \hline
 & \multicolumn{3}{|c|}{Seeley coefficients $a=1.4$} \\
$n$ & $N=1$ & $N=2$ & $N=3$ \\ \hline
1 &  12.1230 & 5.24111 & -1.64075 \\
2 &  36.9804 & 23.4002 & 0.766477 \\
3 &  78.2292 & 62.1502 & 8.35517 \\
4 &  123.711 & 110.113 & 21.0439 \\
5 &  152.580 & 143.635 & 33.2205 \\
6 &  154.368 & 149.554 & 40.3358 \\
7 &  131.749 & 129.557 & 39.4656 \\
8 &  98.6649 & 97.7998 & 33.9219 \\
9 &  60.7463 & 60.4450 & 20.6657 \\
10 & 50.0015 & 49.9076 & 27.8002 \\
11 & -27.4816 & -27.5086 & -38.6083 \\\hline
\end{tabular}}}

The one-instanton determinants $\mathbf{D}(a)$ in units of $(\Omega^2)^{-1}$ are listed in the last Table:

\begin{center}
\begin{tabular}{|c|c|c|c|} \hline
 & $N=1$ & $N=2$ & $N=3$ \\ \hline
$a=0.01$ & $4.02821 \times 10^{-8}$ & $4.0285\times 10^{-8}$ & $4.02816 \times 10^{-8}$ \\ \hline
$a=0.2$ & $0.0000255126$ & $0.000025513$ & $0.0000255125$ \\ \hline
$a=0.4$ & $0.000202913$ & $0.000202912$ & $0.000202909$ \\ \hline
$a=0.6$ & $0.000687855$ & $0.000687854$ & $0.000687846$ \\ \hline
$a=0.8$ & $0.00137759$ & $0.0013776$ & $0.00137559$ \\ \hline
$a=1.0$ & $0.0020107$ & $0.00201088$ & $0.00201067$ \\ \hline
$a=1.2$ & $0.00243354$ & $0.00243354$ & $0.00243351$ \\ \hline
$a=1.4$ & $0.00263335$ & $0.00263336$ & $0.00263333$ \\ \hline
\end{tabular}
\end{center}

\section{Brief outlook}

We look forward to apply these methods in the analysis of the decay of false vacua. In this kind of problems bounces, rather than instantons, lead the semi-classical tunnel effect. Bounce fluctuation operators have typically
one negative mode, their determinants are defined through a process of analytic continuation giving rise to an imaginary energy, a characteristic of resonant states. We believe that our method is very well suited to deal also
with systems of more than one degrees of freedom, a situation considered in the Chapter 11, Section 2, of Reference
\cite{Rubakov}. In this context, for instance, a two dimensional quantum Hamiltonian built from the potential chosen in the scalar field model discussed in \cite{Alonso2002} looks very promising to deal with because it corresponds to
a partially integrable mechanical system such that a manifold of instantons is known. Recently, the improved zeta function procedure used in this paper has been tested precisely in this model from a doamin wall perspective to
compute the one-loop wall tension shifts, see \cite{Alonso2014}.

\vspace{0.3cm}

\begin{theacknowledgments}
JMG warmly thanks to Alexander Andrianov and Domenech Espriu for inviting him to participate in the II High-Energy Physics Russian-Spanish Congress in Sankt Petersburg 2013, JMG is also indebted to Mikhail Braun and Mikhail Ioffe for their kind hospitality in his several visits to the Fock Institute in Peterhoff. Both AAI and JMG acknowledge M. Braun and M. Ioffe for their generous cooperation in the development of quantum and theoretical physics in the Northwest of the Iberian peninsula.
\end{theacknowledgments}

\bibliographystyle{aipproc}   

\begin{thebibliography}{99}

\bibitem{Gelfand}I.M. Gelfand and A.M. Yaglom, {\sl \lq\lq Integration in functional spaces and its applications in quantum mechanics\rq\rq}, Jour. Math. Phys. {\bf 1}(1960) 48

\bibitem{RaySinger} D. B. Ray and I. M. Singer, {\sl \lq\lq R-torsion and the Laplacian on Riemannian manifolds\rq\rq}, Adv. in Math. {\bf 7}:145-210,1971

\bibitem{Gilkey1984} P.B. Gilkey; \textit{Invariance theory, the heat equation and the Atiyah-Singer index theorem}, Publish or Perish, Inc 1984.

\bibitem{DeWitt1965} B.S. de Witt, \textit{Dynamical theory of groups and fields}, Gordon and Breach, 1965.

\bibitem{Elizalde1994} E. Elizalde, S. Odintsov, A. Romeo, A. Bytsenko, S. Zerbini, \textit{Zeta regularization techniques with
applications}, Singapore, World Scientifique, 1994.

    \bibitem{Kirsten2002} D.V. Kirsten, \textit{Spectral functions in
mathematics and physics}, Chapman and Hall/CRC, New York, 2002.

\bibitem{Vassilevich2003} D.V. Vassilevich, {\sl \lq\lq Heat kernel expansion: user's manual\rq\rq}, Phys. Rep. 388C (2003) 279-360.

\bibitem{Alonso2012a} A. Alonso-Izquierdo, J. Mateos Guilarte, {\sl \lq\lq Kink fluctuation asymptotics and zero modes\rq\rq},  Eur. Phys. J. C 72 (2012) 2170.

\bibitem{Alonso2013}A. Alonso-Izquierdo and J. Mateos Guilarte, {\sl \lq\lq Gilkey-DeWitt heat kernel expansion and zero modes\rq\rq}, Il Nuov. Cim. {\bf 36C}(2013) 11519

\bibitem{Polyakov} A. Polyakov, {\sl \lq\lq Quark confinement and the topology of gauge fields\rq\rq}, Nucl. Phys. {\bf B 120}: 429-458, 1977.

\bibitem{Coleman} S. Coleman, {\sl \lq\lq The Uses of Instantons "}, in Aspects of symmetry, Cambridge University
Press, 1985.

\bibitem{Rubakov} V.A. Rubakov, {\sl \lq\lq Classical Theory of Gauge Fields\rq\rq}, Princeton University Press, 2002.

\bibitem{Dunne} G. Dunne, {\sl \lq\lq Functional determinants in quantum field theory\rq\rq}, Jour. Phys. {\bf A41} (2008) 30406

\bibitem{Alonso2004} A. Alonso-Izquierdo, W. Garcia
Fuertes,   M.A. Gonzalez Leon, J.Mateos Guilarte, {\sl \lq\lq One-loop corrections to classical masses of kink families\rq\rq}, Nucl. Phys. {\bf B 681}: 163-194, 2004

\bibitem{Barton} G. Barton, {\sl \lq\lq Levinson's theorem in one dimension: heuristics\rq\rq}, J. Phys. {\bf A18} (1985) 479

\bibitem{Wipf} A. W. Wipf, {\sl \lq\lq Tunnel determinants\rq\rq}, Nucl. Phys. {\bf B269}: 24-44, 1986

\bibitem{Marcos} M. Mari$\tilde{n}$o, {\sl \lq\lq Instantons and large N\rq\rq}, laces.web.cern.ch/Laces10/notes/instlargen.pdf

\bibitem{Mateos2006} A. Alonso-Izquierdo, W. Garc\'{\i}a Fuertes, M. A. Gonzalez Leon, M. de la Torre Mayado, J. Mateos Guilarte, J. M. Mu\~noz-Casta\~neda, {\sl\lq\lq Lectures on the mass of topological solitons"}, arXiv[ hep-th/0611180].

\bibitem{Mateos2009} J. Mateos Guilarte, A. Alonso-Izquierdo, W. Garc\'{\i}a Fuertes, M. de la Torre Mayado, M. J.Senosia\'{\i}n, {\sl \lq\lq Quantum fluctuations around low dimensional topological defects\rq\rq},
Proceedings of Science (ISFTG) 013 (2009) (63pp).

\bibitem{Razavy} M. Razavy, {\sl \lq\lq An exactly solvable Schr$\ddot{\rm o}$dinger equation with a bistable potential\rq\rq}, Amer. Jour. Phys. {\bf 48}(1980) 285

\bibitem{Khare} D.P. Jatkar, C.N. Kumar, A. Khare,  {\sl \lq\lq A quasi-exactly solvable problem without SL(2) symmetry\rq\rq}, Phys. Lett, {\bf A142}(1989)200

\bibitem{Alonso2002} A. Alonso-Izquierdo,  M.A. Gonzalez Leon, J.Mateos Guilarte, {\sl \lq\lq Kink variety in systems of two coupled scalar fields in two space-time dimensions\rq\rq}, Phys. Rev. {\bf D65} (2002) 085012

\bibitem{Alonso2014} A. Alonso-Izquierdo, J.Mateos Guilarte, {\sl \lq\lq Quantum induced interactions in the moduli space of BPS domain walls\rq\rq}, JHEP{\bf  01} (2014) 125

\end{thebibliography}

\end{document}